\documentclass[twoside,a4paper,11pt]{proceedings}
\usepackage{graphicx}
\usepackage{hyperref}
\usepackage{movie15}
\usepackage{natbib}
\begin{document}
\pagenumbering{arabic}
\pagestyle{myheadings}
\thispagestyle{empty}
\vspace*{-1cm}
{\flushleft\includegraphics[width=8cm,viewport=0 -30 200 -20]{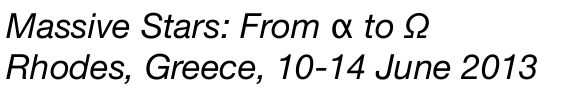}}
\vspace*{0.2cm}
\begin{flushleft}
{\bf {\LARGE
Models of extended decretion disks of critically 
rotating stars
}\\
\vspace*{1cm}
Petr Kurf\"urst$^1$,
Achim Feldmeier$^2$,
and
Ji\v r\'\i\ Krti\v cka$^1$
%
}\\
\vspace*{0.5cm}
%
$^{1}$
\'Ustav teoretick\'e fyziky a astrofyziky, Masarykova univerzita, Brno, Czech Republic \\
$^{2}$
Institut f\"ur Physik und Astronomie, Universit\"at Potsdam, Potsdam-Golm, Germany 
\end{flushleft}
\markboth{
Models of extended decretion disks 
}{
Petr Kurf\"urst et al.
}
\thispagestyle{empty}
\vspace*{0.4cm}
\begin{minipage}[l]{0.09\textwidth}
\ 
\end{minipage}
\begin{minipage}[r]{0.9\textwidth}
\vspace{1cm}
\section*{Abstract}{\small
Massive stars may during their 
evolution reach the phase of critical rotation when 
the further increase in rotational speed is no longer possible.
The ejection of matter in the equatorial region forms the gaseous outflowing disk, 
which allows the star to remove the excess of its angular momentum. The outer part of 
the disk can extend up to large distance from the parent star. 
We study the evolution of density, radial and azimuthal velocity and angular momentum 
loss rate of equatorial decretion disk up to quite distant outer regions. 
We calculate the evolution of density, radial and azimuthal velocity 
from the initial Keplerian state until the disk 
approaches the final stationary state. 
\vspace{10mm}
\normalsize}
\end{minipage}

\section{Viscous decretion disks of critically rotating stars}
 
The basic scenario follows the model of the viscous decretion disk proposed by 
\citet{Lee}, which naturally leads to the formation of 
Keplerian disk, assuming the outward transport of matter from star's near-equatorial surface occurs through 
the gradual drifting and feeding of the disk due to the viscous torque. 
The disk mass loss rate of critically rotating star is determined by the 
angular momentum loss rate needed to keep the star at critical rotation 
\citep{Krticka}.
The main uncertainties are the viscous coupling and the radial temperature distribution. 
The modelling of the viscosity is 
typically constrained to regions close to the star \citep{Penna}, 
consequently we assume a power low decrease of the viscous coupling. In this model we maintain 
the temperature distribution as a free parameter.
Following the stationary models \citep{Kurf} we introduce the results of calculations of the time-evolving 
converging solutions for governing physical quantities
in the disk up to quite distant supersonic regions from the parent star.

\section{Radial thin disk structure}

Basic hydrodynamic equations (mass, momentum and energy conservation equations) determine the disk behaviour. 
We assume axial symmetry in cylindrical coordinate system and fully integrated vertical structure 
with the vertically integrated density $\Sigma=\int_{-\infty}^{\infty}\rho\,\rm{d}z$ and with 
constant temperature and hydrostatic equilibrium in $z$-direction \citep{Lee, Krticka, Kurf}.
Disk models \citep{Carcio} found nearly isothermal 
($T_0\approx\frac{1}{2}T_{\rm{eff}}$) temperature distribution in the inner disk region.  
For calculations of the structure of outer part of the disk it is reasonable to consider the power law temperature decline
$T=T_0(R_{\rm{eq}}/R)^p$, where $R$ is the cylindrical radial distance and $p$ is a free parameter.
In our models we employ the second order Navier-Stokes viscosity.
\citet{Penna} find the nonrelativistic $\alpha$ viscosity coefficient being constant
$\alpha=0.025$ close to the star.
We examine also disk behaviour with the introduced 
power law viscosity dependence $\alpha=\alpha_0(R_{\rm{eq}}/R)^n$, where $\alpha$ is 
the viscosity parameter \citep{Shak}, $n$ is a free parameter and $\alpha_0$ denotes viscosity near the stellar surface.

\section{Results of numerical solution}

\begin{figure}[h!]
\center
\includegraphics[width=0.42\textwidth]{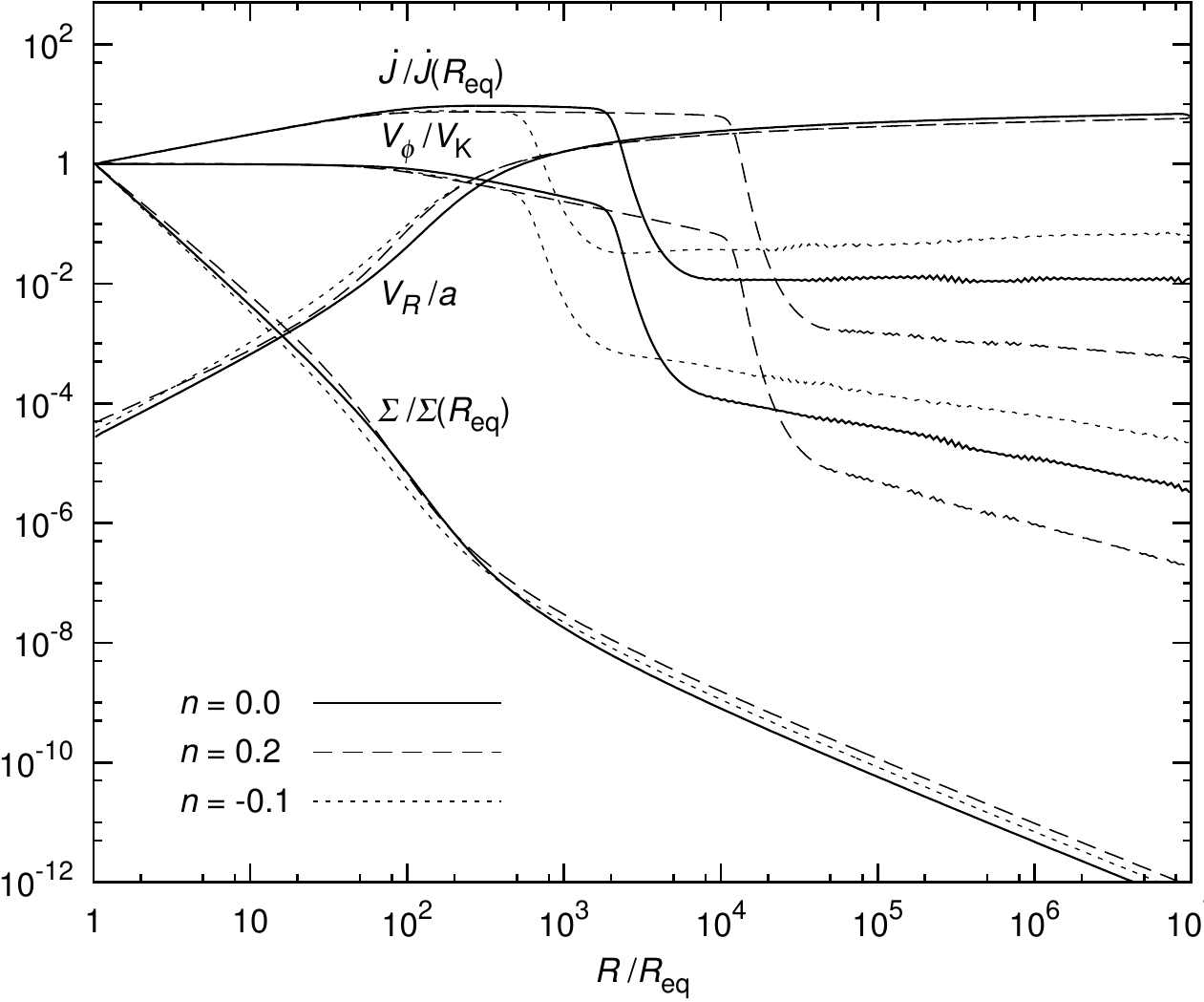}~~~~~~~~\includegraphics[width=0.42\textwidth]{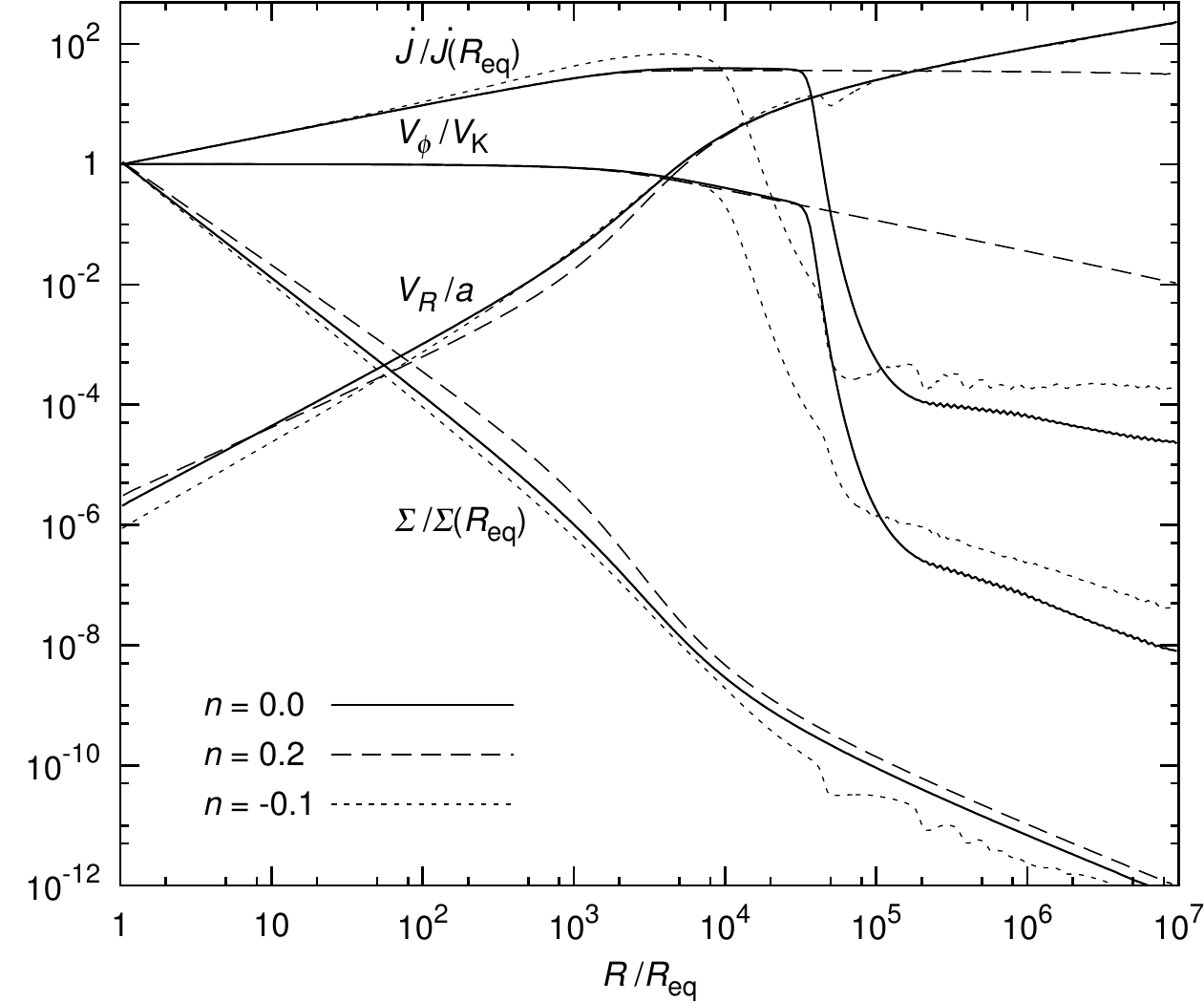} 
\caption{The dependence of the relative disk column density ($\Sigma/\Sigma(R_{\rm{eq}})$),
the relative radial ($V_R/a$) and azimuthal ($V_{\phi}/V_{\rm{K}}$) disk velocities
(in the units of sound speed $a$ and Keplerian velocity $V_{\rm{K}}$)
and the relative angular momentum  loss rate ($\dot{J}/\dot{J}(R_{\rm{eq}})$)
on radius in case of isothermal disk (left graph) and of the power law temperature decrease with parameter $p=0.4$ 
(right graph) in final stationary models.
Viscosity profiles are represented by parameter ${n}$ in the graph, 
inner boundary viscosity ${\alpha_0=0.025}$ is considered. The unphysical drop of the angular momentum loss rate at large radii 
can be avoided with decreasing viscosity coefficient  $\alpha$.
In this case the disks become angular momentum conserving at large radii.}
\label{graph1}
\end{figure}

\begin{figure}[h!]
\center
\includegraphics[height=5.85cm]{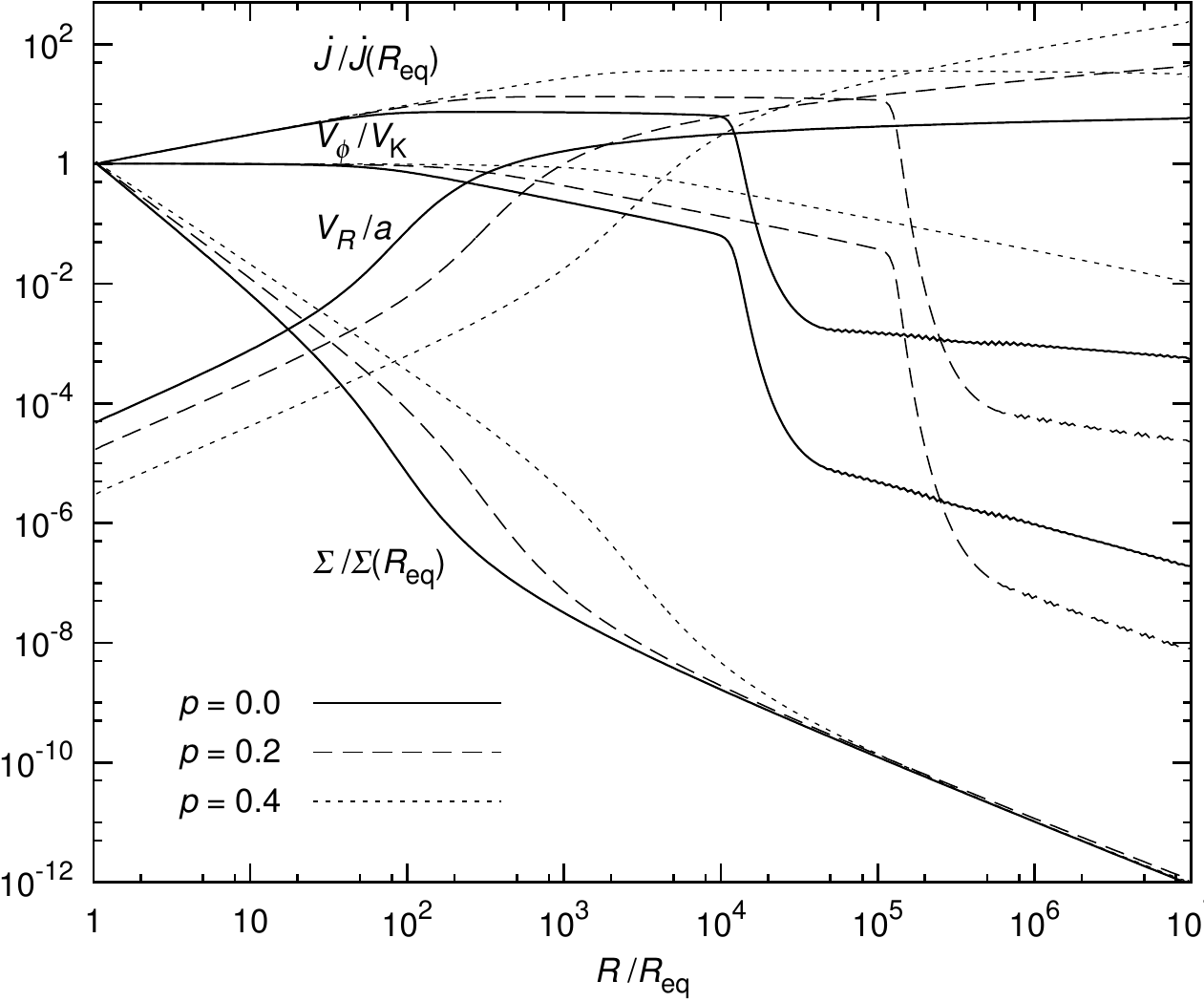}~~~~~~~~~~\includegraphics[height=6cm]{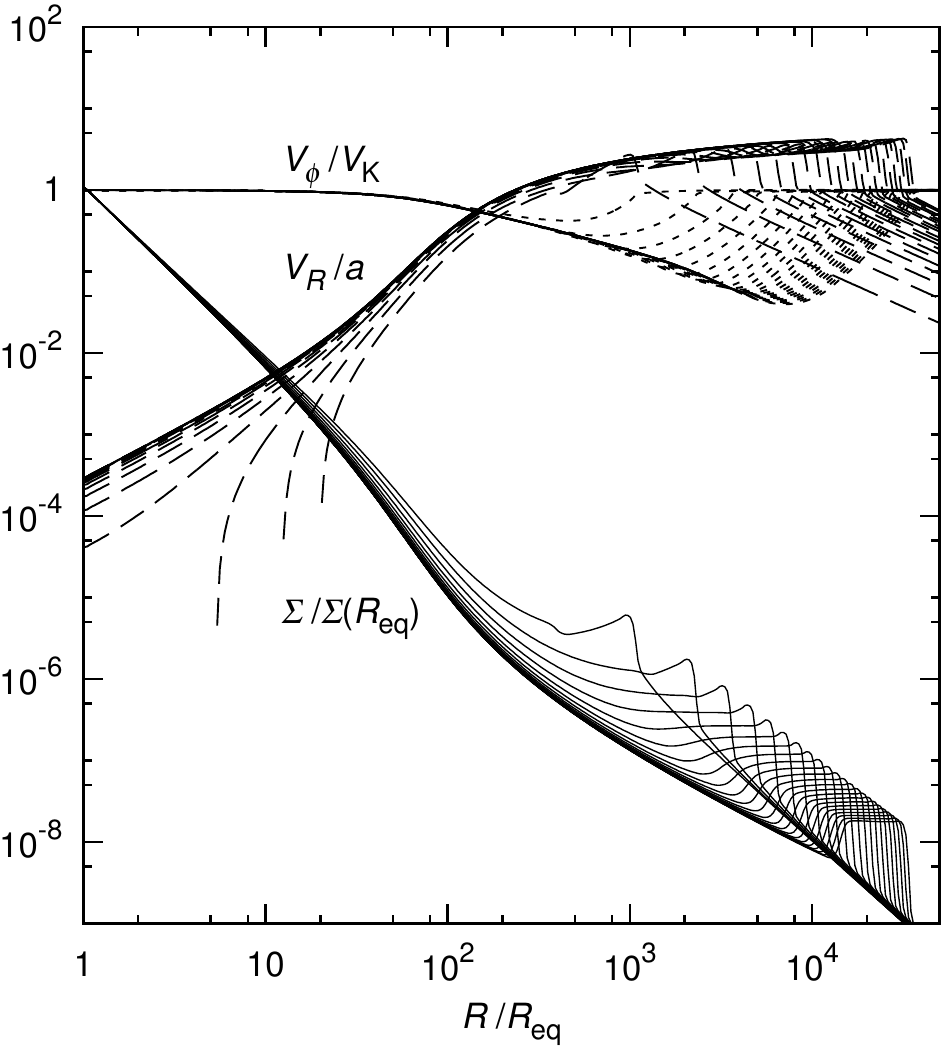} 
\caption{Left panel: The same as Fig. \ref{graph1}, however with 
decreasing viscosity $\alpha=\alpha_0(R_{\rm{eq}}/R)^{0.2}$ for different temperature profiles 
parameterized by $p$. 
Right panel: 
Snaphsots of the density and velocity profiles plotted each 
5.1 years of the disk evolution. The graph shows the density wave 
propagation on the viscous time scale $t_{\rm{visc}}\sim R^2/\nu$, which can be considered as as 
the timescale for a disk annulus to move a radial distance $R$ \citep{Pringle}.} 
\label{graph2}
\end{figure}

For the purpose of modelling we developed the time-depedent hydrodynamic grid code.
As a central object we selected the B0 type star with the following parameters \citep{Harma}:
$T_{\mathrm{eff}}=30\,000\,{\rm{K}},\,M=14.5\,{\rm{M}}_{\odot},\,R=5.8\,{\rm{R}}_{\odot}.$
The results of our calculations are given in Figs.~\ref{graph1},~\ref{graph2}.

\small  
%
\section*{Acknowledgments}   
%
This work was supported by the grant GA \v{C}R 13-10589S.

\bibliographystyle{aj}
\small
\bibliography{proceedings}

\end{document}